\documentclass[twocolumn,showpacs,amsmath,amssymb,prb,floatfix,aps]{revtex4-1}
\usepackage{graphicx}
\usepackage{dcolumn}
\usepackage{bm}
\begin{document}
\title{
Ferrimagnetism 
in the Spin-1/2 Heisenberg Antiferromagnet 
on a Distorted Triangular Lattice 
}

\author{Hiroki Nakano$^{1}$, and T\^oru Sakai$^{2}$}
\affiliation
{$^{1}$
Department of Material Science, Graduate School of Material Science, 
University of Hyogo, \\
3-2-1 Kouto, Kamigori, Hyogo, 678-1297, Japan \\
$^{2}$
National Institutes for Quantum and Radiological Science and Technology 
(QST),
SPring-8
Sayo, Hyogo 679-5148, Japan 
}
\date{\today}

\begin{abstract}
The ground state of the spin-$1/2$ Heisenberg antiferromagnet 
on a distorted triangular lattice is studied
using a numerical-diagonalization method. 
The network of interactions is the $\sqrt{3}\times\sqrt{3}$ type;
the interactions are continuously controlled between
the undistorted triangular lattice and the dice lattice.
We find new states between the nonmagnetic
120-degree-structured state of the undistorted triangular case 
and the up-up-down state of the dice case.
The intermediate states show spontaneous magnetizations 
that are smaller than one third of the saturated magntization 
corresponding to the up-up-down state. 
\end{abstract}

\maketitle

Frustration is becoming
considerably more relevant 
in modern condensed-matter physics 
because nontrivial and fascinating phenomena
often occur owing to the quantum nature 
in systems including frustrations. 
A typical frustrated magnet 
is the triangular-lattice antiferromagnet. 
This system includes regular triangles 
composed of bonds of antiferromagnetic interaction 
between nearest-neighbor spins. 
A few decades ago,
the quantum Heisenberg antiferromagnet on the triangular lattice 
became one of the central issues 
as a candidate system of the spin liquid state\cite{Anderson_tri}; 
extensive studies have long been carried out\cite{Huse_Elser,
Jolicour_LGuillou,Singh_Huse,Bernu1992,Bernu1994,Leung_Runge,
Richter_lecture2004,Sakai_HN_PRBR,DYamamoto2013}.
Further, good experimental realizations were
reported\cite{Shirata_S1tri_2011,Shirata_Shalf_PRL2012}.
Distortion effects along one direction\cite{Weihong1999,Yunoki2006,
Starykh2007,Heidarian2009,Reuther2011,Weichselbaum2011,Ghamari2011,
Harada2011,s1tri_LRO} and
randomness effects\cite{KWatanabe_JPSJ2015,Shimokawa_JPSJ2015} have
also been investigated. 

Recently, it is widely believed that
the ground state of the triangular-lattice antiferromagnet reveals
a spin-ordered state of the so-called 120-degree structure 
without a magnetic field.  
If a magnetic field is applied to this system, 
it is well known that a magnetization plateau appears
at one-third of the saturation magnetization
in the zero-temperature magnetization curve, 
although the corresponding classical case does not show this plateau.
The spins at the plateau are considered to be collinear,
and the state is called the up-up-down state. 
In the magnetization curves, this system shows
a Y-shaped spin state between
the 120-degree-structured state under no field, 
and the up-up-down state in the magnetization plateau
under a field. 

On the other hand, a similar up-up-down state
is also realized in the spin model on the dice lattice
as its ground state without a magnetic field\cite{Dice_Jagannathan}. 
The dice lattice is a bipertite one;
frustration disappears and
the Lieb-Mattis (LM) theorem holds\cite{Lieb_Mattis}. 
Therefore,
the up-up-down state of this model is the ferrimagnetic one
based on the LM theorem. 
It is worth emphasizing that
this state originates only from the lattice structure
in spite of the fact that a field is not applied. 
Note additionally that the dice lattice is obtained 
by the removal of parts of the interaction bonds
in the triangular lattice. 

Under these circumstances, we are faced with a question:
what is the spin state in the Heisenberg antiferromagnet 
when one continuously controls the interaction bonds 
between triangular and dice lattices?
Note here that the controlling of the bonds corresponds
to the $\sqrt{3}\times\sqrt{3}$ distortion in the triangular lattice. 
The purpose of this letter is to clarify the behavior
of the change in the spin state
in the $\sqrt{3}\times\sqrt{3}$-distorted triangular lattice. 
The present numerical-diagonalization results
provide us with a new route to change a spin state
from the 120-degree-structured state 
to the up-up-down spin state without applying a magnetic field. 

\begin{figure}[b]
%
\begin{center}
\includegraphics[width=0.34\textwidth]{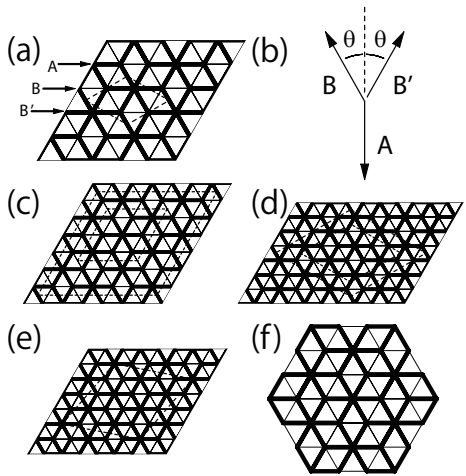} 
%
\end{center}
\caption{
Distorted triangular lattice and finite-size clusters.
Panel (a) illustrates sites A, B, and B$^{\prime}$ as well as the unit cell
of the system. Panel (b) depicts the classical picture.
The rhombuses in panel (c) illustrate finite-size clusters for $N_{\rm s}=9$ and 36
under the periodic boundary condition; 
rhombuses in panel (d) for $N_{\rm s}=12$ and 27;  
rhombus in panel (e) for $N_{\rm s}=21$.
Panel (f) illustrates the $N_{\rm s}=37$ cluster
under the open boundary condition. 
}
\label{fig1}
\end{figure}

The Hamiltonian studied in this letter is given by 
\begin{eqnarray}
{\cal H} &=& 
\sum_{i \in \mbox{B},j \in \mbox{B}^{\prime}} J_{1} 
\mbox{\boldmath $S$}_{i}\cdot\mbox{\boldmath $S$}_{j} 
\nonumber \\ 
& & 
+\sum_{i \in \mbox{A},j \in \mbox{B}} J_{2} 
\mbox{\boldmath $S$}_{i}\cdot\mbox{\boldmath $S$}_{j} 
+\sum_{i \in \mbox{A},j \in \mbox{B}^{\prime}} J_{2} 
\mbox{\boldmath $S$}_{i}\cdot\mbox{\boldmath $S$}_{j} 
. 
\label{Hamiltonian}
\end{eqnarray}
Here, $\mbox{\boldmath $S$}_{i}$ 
denotes the $S=1/2$ spin operator at site $i$. 
In this study, we consider 
the case of isotropic interaction in spin space. 
The site $i$ is assumed to be the vertices 
of the lattice illustrated in Fig.~\ref{fig1}. 
The number of spin sites is denoted by $N_{\rm s}$.  
The vertices are divided into three sublattices 
A, B, and B$^{\prime}$; 
each site $i$ in the A sublattice is linked 
by six interaction bonds $J_{2}$ denoted by thick lines; 
each site $i$ in the B or B$^{\prime}$ sublattice is linked 
by three interaction bonds $J_{2}$ and 
three interaction bonds $J_{1}$, denoted by thin lines. 
We denote the ratio of $J_{2}/J_{1}$ by $r$.  
We consider that all interactions are antiferromagnetic, 
namely, $J_{1} > 0$ and $J_{2} > 0$. 
Energies are measured in units of $J_{1}$; 
hereafter, we set $J_{1}=1$ 
and examine the case of $J_{2} \ge J_{1}$. 
Note that for $J_{1}=J_{2}$, namely, $r=1$, 
the present lattice is identical to the triangular lattice, 
where the ground state is well known as a nonmagnetic state. 
For $J_{1}\rightarrow 0$, namely, $r\rightarrow\infty$, 
on the other hand, the network of the vertices becomes the dice lattice. 

The finite-size clusters that we treat in the present study 
are depicted in Fig.~\ref{fig1}(c)-(f).  
We examine the cases of $N_{\rm s}=9$, 12, 21, 27, and 36 
under the periodic boundary condition 
and the case of $N_{\rm s}= 37$ 
under the open boundary condition. 
In the former cases, $N_{\rm s}/3$ is an integer; 
therefore, the number of spin sites in a sublattice 
is the same irrespective of sublattices. 
The clusters in the former cases 
are rhombic and have an inner angle $\pi/3$; 
this shape allows us to capture two dimensionality well. 

We calculate the lowest energy of ${\cal H}$ 
in the subspace belonging to $\sum _j S_j^z=M$ 
by numerical diagonalizations 
based on the Lanczos algorithm and/or the Householder algorithm. 
The numerical-diagonalization calculations 
are unbiased against any approximations;
one can therefore obtain reliable information of the system. 
The energy is denoted by $E(N_{\rm s},M)$, 
where $M$ takes an integer or a half odd integer up 
to the saturation value $M_{\rm sat}$ ($=N_{\rm s}/2$). 
We define $M_{\rm spo}$ as the largest value of 
$M$ among the lowest-energy states, 
because we focus our attention on spontaneous magnetization. 
Note, first, that for cases of odd $N_{\rm s}$, 
the smallest $M_{\rm spo}$ cannot vanish; 
the result of $M_{\rm spo}=1/2$ in the ground state indicates 
that the system is nonmagnetic. 
We also use the normalized magnetization $m=M_{\rm spo}/M_{\rm sat}$. 
Part of the Lanczos diagonalizations were carried out 
using a MPI-parallelized code, which was originally 
developed in the study of Haldane gaps\cite{HN_Terai}. 
The usefulness of our program was confirmed in large-scale 
parallelized calculations\cite{kgm_gap,s1tri_LRO,
HN_TSakai_kgm_1_3,HN_TSakai_kgm_S,HN_YHasegawa_TSakai_dist_shuriken}. 

%

Before observing our diagonalization results for the quantum case, 
let us consider the classical case composed of classical vectors 
with length $S$.  
A probable spin state is depicted in Fig.~\ref{fig1}(b). 
For a given $r$,  
minimizing the energy determines the angle $\theta$ related to $M_{\rm spo}$. 
One obtains $M_{\rm spo}/M_{\rm sat}=1/3$ for $r \ge 2$ 
and $M_{\rm spo}/M_{\rm sat}=(r-1)/3$ for $1 < r \le 2$.  
The same spin state was discussed in ref.~\ref{Nishiwaki_jpsj2011}. 


%
\begin{figure}[b]
%
\begin{center}
\includegraphics[width=0.34\textwidth]{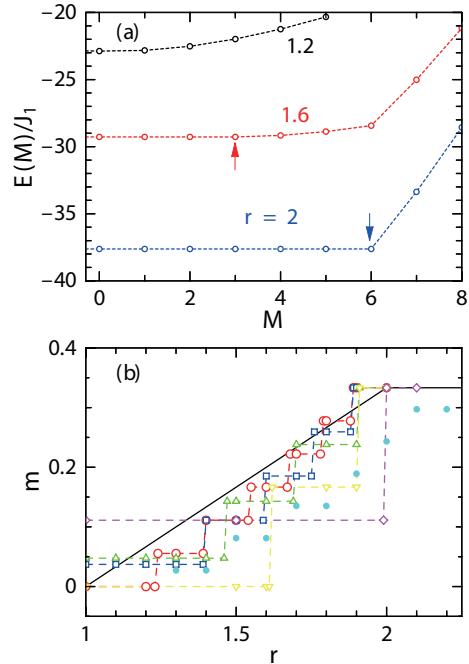} 
%
\end{center}
\caption{(Color) 
(a) $M$-dependence of the ground-state energy for $N_{\rm s}=36$. 
(b) $r$-dependence of the spontaneous magnetization 
for various system sizes.
Violet diamonds, yellow reversed triangles, 
green triangles, blue squares, and red circles denote results
for $N_{\rm s}=9$, 12, 21, 27, and 36 under the periodic boundary condition.
Light blue closed circles denote results for $N_{\rm s}=37$ 
under the open boundary condition.
}
\label{fig2}
\end{figure}

%
\begin{figure}[b]
%
\begin{center}
\includegraphics[width=0.34\textwidth]{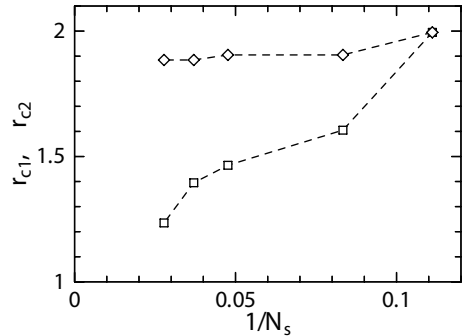} 
%
\end{center}
\caption{
System-size dependence of the critical ratios 
for various system sizes.
The squares and diamonds denote the results for $r_{\rm c1}$ and $r_{\rm c2}$,
respectively. 
}
\label{fig3}
\end{figure}

Now, we present our results for the quantum case. 
First, our data for the lowest energy 
in each subspace of $M$ is shown in Fig.~\ref{fig2}(a), 
which depicts the case for $N_{\rm s}=36$.  
This figure reveals whether a spontaneous magnetization 
occurs, and its magnitude if it occurs. 
For $r=1.2$, the energy for $M=0$ is lower 
than the energies for a larger $M$, which indicates that
the ground state is nonmagnetic. 
For $r=1.6$, on the other hand, 
the energies for $M=0$ to 3 are numerically identical, 
which means that the system shows a spontaneous magnetization,  
and this magnetization is $M_{\rm spo}=3$. 
For $r=2$, 
the energies for $M=0$ to 6 are numerically identical,
and the spontaneous magnetization $M_{\rm spo}=6$. 
Since the saturation is $M_{\rm sat}=18$ in the $N_{\rm s}=36$ system, 
$m=1/3$ suggests that 
the LM ferrimagnetic state is realized. 
Figure~\ref{fig2}(a) strongly suggests that 
the present system shows gradual magnetization 
owing to the distortion $r > 1$
between the nonmagnetic state and the LM ferrimagnetic state. 
These intermediate states may be interpreted as a collapse
of ferrimagnetism occurring in the dice-lattice antiferromagnet. 
Note that we have so far investigated a collapse
of ferrimagnetism occurring in the Lieb-lattice antiferromagnet
by various distortions. 
The distorted kagome-lattice 
antiferromagnet shows a similar intermediate
state\cite{collapse_ferri2d,Shimokawa_JPSJ}, which
will be compared to later. 
In other various distortions, 
such intermediate states were not detected so far\cite{shuriken_lett,
HN_kgm_dist,HNakano_Cairo_lt,Isoda_Cairo_full,HN_TSakai_JJAP_RC}.

Next, we examine the intermediate state in detail. 
Our results are depicted in Fig.~\ref{fig2}(b). 
For $N_{\rm s}=9$, 
the nonmagnetic state of $M_{\rm spo}=1/2$ 
and 
the LM ferrimagnetic state of $M_{\rm spo}=3/2=(1/3)M_{\rm sat} $ 
are neighboring with each other without an intermediate $M_{\rm spo}$; 
however, this behavior comes from the smallness of $N_{\rm s}$. 
For a larger $N_{\rm s}$, there appear intermediate-$M_{\rm spo}$ 
states between the nonmagnetic state of $M_{\rm spo}=0$ or 1/2 and  
the LM ferrimagnetic state of $M_{\rm spo}=(1/3)M_{\rm sat}$. 
Note here that the states of all possible $M_{\rm spo}$ are realized 
between the smallest $M_{\rm spo}$ and $(1/3)M_{\rm sat}$ 
irrespective of $N_{\rm s} \ge 12$. 
Another marked behavior is that 
the range of the ratio $r$ of the intermediate $M_{\rm spo}$ 
gradually widens 
as $N_{\rm s}$ is increased. 
In order to clarify this behavior, 
we plot the $N_{\rm s}$-dependences of the critical ratios 
depicted in Fig.~\ref{fig3}. 
Here, we define $r_{\rm c1}$ as the value of $r$ 
at which the ground state 
changes from $M_{\rm spo}=0$ or 1/2 to the next $M_{\rm spo}$, 
and $r_{\rm c2}$ as the value of $r$ at which the ground state 
changes to $M_{\rm spo}=(1/3)M_{\rm sat}$ from $M_{\rm spo}=(1/3)M_{\rm sat}-1$. 
Note that for $N_{\rm s}=9$, $r_{\rm c1}=r_{\rm c2}$ as mentioned above. 
One can easily observe that $r_{\rm c2}$ shows 
a  very weak system size dependence. 
It is expected that an extrapolated value is $r_{\rm c2} \sim 1.9$. 
On the other hand, $r_{\rm c1}$ gradually decreases 
as $N_{\rm s}$ is increased. 
Although the dependence is not smooth, 
our numerical data suggest
that an extrapolated value of $r_{\rm c1}$ is very close 
to $r=1$ corresponding to the case of the undistorted triangular lattice. 
It is difficult to determine $\lim_{N_{\rm s}\rightarrow \infty} r_{\rm c1} $ 
precisely from the present samples of finite sizes. 
To determine a reliable consequence concerning 
whether this limit is equal to 1 or different from 1, 
further investigations will be required. 
Note that even if this limit equals 1, 
such a consequence is consistent with the modern understanding of the 
triangular-lattice antiferromagnet, revealing 
the 120-degree spin structure in the ground state. 

Let us compare the $r$-dependence of $M_{\rm spo}$ with the classical case,
shown in Fig.~\ref{fig2}(b). 
Our numerical data under the periodic boundary condition 
agree well with the solid line in the classical case 
illustrated in Fig.~\ref{fig1}(b). 
This agreement seems to suggest that the intermediate-$M_{\rm spo}$ 
spin states in the quantum case can be understood 
based on the classical picture. In the following, 
let us examine whether 
this classical picture is valid in the quantum case 
from the viewpoint of the local magnetization $\langle S_{i}^{z}\rangle$. 
Here, the symbol $\langle {\cal O} \rangle$ 
represents the expectation value of the operator ${\cal O}$ 
with respect to the lowest-energy state within the subspace 
characterized by a magnetization $M_{\rm spo}$.

%
\begin{figure}[b]
%
\begin{center}
\includegraphics[width=0.34\textwidth]{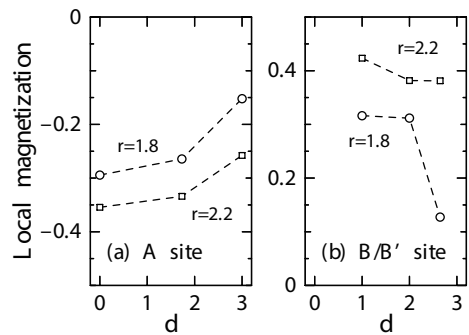} 
%
\end{center}
\caption{
Site-dependence of the local magnetizations $\langle S_{i}^{z}\rangle$
for the system of $N_{\rm s}=37$. 
Distance $d$ is the distance between site $i$ and 
the center of the system with the open boundary,
measured in units of the distance 
between two nearest-neighbor sites 
in the triangular lattice for $r=1$. 
}
\label{fig4}
\end{figure}

Figure \ref{fig4} depicts the site-dependence 
of the local magnetizations for the system of $N_{\rm s}=37$ 
under the open boundary condition.  
Owing to this boundary condition, spin sites in each sublattice
are not equivalent.
The sites are divided into groups of equivalent sites
characterized by the distance from the center of the cluster. 
Thus, we present our results  in Fig.~\ref{fig4}
as a function of the distance. 
We study the results for the case 
under the open boundary condition
to compare a similar intermediate-$M_{\rm spo}$ state 
reported in the distorted kagome-lattice 
antiferromagnet\cite{collapse_ferri2d,Shimokawa_JPSJ}, 
in which the local magnetizations 
show a nontrivial incommensurate modulation 
suggesting non-Lieb-Mattis ferrimagnetism. 
Note that 
the realizations of such incommensurate-modulation states 
were originally reported in several one-dimensional systems
\cite{Ivanov_Richter,Yoshikawa_Miya_JPSJ2005,Hida_JPSJ2007,
Hida_JPCM2007,Hida_Takano_PRB2008,Montenegro_Coutinho_PRB2008,
Hida_Takano_Suzuki_JPSJ2010,Shimokawa_Nakano_JPSJ2011,
Furuya_Giamarchi_PRB2014,Hida_JPSJ2016}. 
Therefore, in this study,
we focus on identifying the relationship
between these incommensurate-modulation 
states and the intermediate states. 
We confirm that the intermediate-$M_{\rm spo}$ states appear 
in the case under the open boundary condition 
as depicted in the results in Fig.~\ref{fig2}(b).  
Note that 
$m$ corresponding to the LM ferrimagnetism does not agree with 1/3 
when $N_{\rm s}/3$ is not an integer. 
For $r=2.2$ in the present system, 
no significant behavior corresponding to incommensurate modulation 
is observed in our numerical data away from the boundary 
of the cluster when one excludes results on the open boundary,  
although a small boundary effect penetrates 
into the inside of the cluster. 
This is consistent with the fact that in this case, 
the LM ferrimagnetic state is realized. 
For $r=1.8$ in the present system, on the other hand, 
an intermediate-$M_{\rm spo}$ state appears. 
Even in such a state, 
our numerical data away from the boundary of the cluster 
do not show the behavior of incommensurate modulation. 
Therefore, our present results 
do not detect a direct evidence of the intermediate-$M_{\rm spo}$ state 
in the present system showing non-Lieb-Mattis ferrimagnetism. 
However, future studies under open boundary conditions 
should be carried out to clarify the relationship 
between these incommensurate-modulation states 
and the intermediate states in the present study.

%
\begin{figure}[b]
%
\begin{center}
\includegraphics[width=0.34\textwidth]{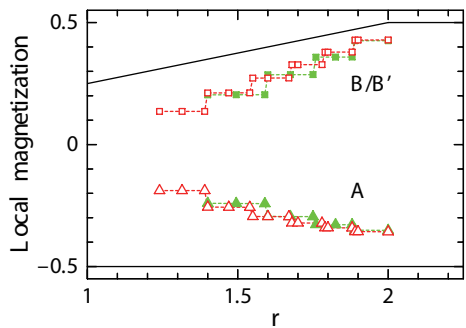} 
%
\end{center}
\caption{(Color) 
$r$-dependence of local magnetizations for 
$r\gtrsim r_{\rm c1}$. 
Solid lines represent the local magnetization from the classical picture. 
Squares denote the numerical results for the local magnetizations 
of B or B$^{\prime}$ sites; triangles denote those for A sites. 
Green closed symbols and red open symbols are for $N_{\rm s}=27$ and 36, 
respectively. 
}
\label{fig5}
\end{figure}

 Next, we examine the $r$-dependence 
of the local magnetizations 
under the periodic boundary condition.
Our numerical results for $N_{\rm s}=27$ and 36  
in the region of $r\gtrsim r_{\rm c1}$ 
are depicted in Fig.~\ref{fig5}. 
Note first that 
under this boundary condition, all spin sites in each sublattice
are equivalent.
Therefore, the numerical results of $\langle S_{i}^{z} \rangle$
at equivalent sites agree with each other within numerical errors.
This is why 
how to present the numerical results of $\langle S_{i}^{z} \rangle$ 
is different between Figs.~\ref{fig4} and \ref{fig5}. 
A significant feature in Fig.~\ref{fig5} is that
the results from the two sizes agree well with each other 
although our numerical results show step-like behaviors, 
which originate from a finite-size effect, as in Fig.~\ref{fig2}(b). 
Owing to this serious finite-size effect, 
it is quite difficult 
to extrapolate finite-size $\langle S_{i}^{z} \rangle$ 
for a given fixed $r$ 
to the limit of $N_{\rm s}\rightarrow\infty$. 
Let us consider 
why we limit the region to be $r \gtrsim r_{\rm c1}$, 
in which our date are presented in Fig.~\ref{fig5}. 
In this region, spontaneous magnetization occurs; therefore, 
the $z$-axis has a specific role, 
which means that the examination of the local magnetizations 
in Fig.~\ref{fig5} contributes considerably to our understanding 
of the magnetic structure of the intermediate states.  
If the spontaneously magnetized intermediate states show 
the same structure as the Y-shaped classical one, 
A-sublattice spins are supposed to be antiparallel to external field 
and B/B$^{\prime}$-sublattice spins are supposed to have components 
that are opposite to the A-sublattice spins. 
In Fig.~\ref{fig5},  
the lines from the Y-shaped classical picture are also illustrated. 
A marked behavior in the classical picture is that 
the down spin at an A-sublattice site maintains 
its local magnetization to be $-1/2$. 
On the other hand, 
in our numerical data, 
the local magnetization of an A-sublattice spin 
for the quantum systems 
gradually increase 
when $r$ is decreased from $r=2$ to $r=r_{\rm c1}$. 
Certainly, we have to be careful when the comparison is carried out 
between the classical system and 
the present behaviors of the finite-size quantum systems. 
This causes the mixing effect of magnetized states from a quantum nature 
just at $r=1$, 
which makes it difficult to detect a difference 
in the local magnetizations between the classical and quantum cases. 
However, for $r \gtrsim r_{\rm c1}$ 
away from $r=1$; 
such a difficulty does not occurs.  
An important difference of the quantum case from the classical one 
is that the local magnetization of an A-sublattice spin 
shows a significant dependence on $r$ in the quantum case. 
Generally speaking, 
there are two sources for changes of the local magnetization; 
one is a deviation of the spin amplitude and 
the other is the spin angle 
measured from the axis of the spontaneous magnetization.  
In the Y-shaped spin state within the classical picture, 
an A-sublattice spin is antiparallel to the axis 
of the spontaneous magnetization; namely, the spin angle vanishes. 
Recall here the spin amplitude of the 
120-degree structure of the undistorted-triangular-lattice antiferromagnet 
from the spin-wave theory\cite{Jolicour_LGuillou}, 
namely $\langle S_{z} \rangle = 0.239$. 
In the present cases for $r\gtrsim r_{\rm c1}$, 
we are approaching the unfrustrated case of the dice lattice; 
therefore the possibility 
that the spin amplitude becomes smaller 
than $\langle S_{z} \rangle = 0.239$ is quite low. 
Under this circumstance, we focus our attention 
on the numerical results of $N_{s}=36$ around $r=1.3$; 
we have $| \langle S_{i}^{z} \rangle | \sim 0.19$ 
for the A-sublattice spin. 
Such a small value of $| \langle S_{i}^{z} \rangle |$ 
cannot be explained 
only from the deviated spin amplitude 
without a change of the spin angle.  
Therefore, 
the spin state around $r=1.3$ of the quantum case 
is considered to be different 
from the Y-shaped spin state within the classical picture. 
The present analysis of the spin structure in the intermediate state 
is only the first step. 
For a deeper understanding of the spin structure, 
future investigations including 
a two-point correlation function and a chirality 
of the intermediate state are necessary. 

Finally, we would like to comment on the experimental situation. 
Tanaka and Kakurai reported magnetic phase transitions of RbVBr$_{3}$, 
which shows a structure of a distorted triangular lattice, 
although the ratio of the interactions is consequently considered 
to be $r<1$\cite{Tanaka_Kakurai}. 
Nishiwaki {\it et al}. studied RbFeBr$_{3}$, 
which also shows $r<1$\cite{Nishiwaki_jpsj2011}. 
A discovery of a new material with $1< r < 2$ would give 
useful information concerning the intermediate state of ferrimagnetism 
from experiments. 


In summary, we investigated the ground-state properties 
of the spin-$1/2$ Heisenberg antiferromagnet on the triangular lattice 
with a distortion 
by the numerical-diagonalization method. 
Under the conditions that 
the undistorted case is common with 
the triangular-lattice antiferromagnet without a magnetic field, 
and that 
the same up-up-down spin state is commonly realized 
both in the distorted case of the present model 
and in the $m$=1/3-plateau state 
of the triangular-lattice antiferromagnet under a magnetic field,  
we find that 
spontaneous magnetization grows along a new route 
to the $m$=1/3 up-up-down state due to the distortion of the lattice, 
which is different from the well-known route in the magnetization 
process of the triangular-lattice antiferromagnet. 
We are now examining this new state 
with intermediate spontaneous magnetization in more detail; 
the results will be published elsewhere.

We wish to thank 
Professors 
H.~Sato, K.~Yoshimura, N. Todoroki, 
and 
Miss A. Shimada  
for fruitful discussions.
We wish to thank
Dr. James Harries 
for his critical reading of our manuscript. 
This work was partly supported 
by JSPS KAKENHI Grant Numbers 
16K05418, 16K05419, and 16H01080(JPhysics). 
Nonhybrid thread-parallel calculations
in numerical diagonalizations were based on TITPACK version 2
coded by H. Nishimori. 
Some of the computations were 
performed using facilities of 
the Department of Simulation Science, 
National Institute for Fusion Science; 
Institute for Solid State Physics, The University of Tokyo;  
and Supercomputing Division, 
Information Technology Center, The University of Tokyo. 
This work was partly supported 
by the Strategic Programs for Innovative Research; 
the Ministry of Education, Culture, Sports, Science 
and Technology of Japan; 
and the Computational Materials Science Initiative, Japan.

\end{document}